# Pressure-induced melting of magnetic order and emergence of new quantum state in α-RuCl₃

Zhe Wang[1]*, Jing Guo[1]*, F. F. Tafti[2], Anthony Hegg[3], Sudeshna Sen[3], Vladimir A Sidorov[4], Le Wang[1], Shu Cai[1], Wei Yi[1], Yazhou Zhou[1], Honghong Wang[1], Shan Zhang[1], Ke Yang[5], Aiguo Li[5], Xiaodong Li[6], Yanchun Li[6], Jing Liu[6], Youguo Shi[1], Wei Ku[3,7]†, Qi Wu[1], Robert J Cava[2], Liling Sun[1,8]†

[1]*Institute of Physics and University of Chinese Academy of Sciences, Beijing 100190, China*

[2]*Department of Chemistry, Princeton University, Princeton, New Jersey 08544, USA*

[3]*School of Physics and Astronomy, Shanghai Jiaotong University, Shanghai 200240*

[4]*Institute for High Pressure Physics, Russian Academy of Sciences, 142190 Troitsk, Moscow, Russia*

[5]*Shanghai Synchrotron Radiation Facilities, Shanghai Institute of Applied Physics, Chinese Academy of Sciences, Shanghai 201204, China*

[6]*Institute of High Energy Physics, Chinese Academy of Sciences, Beijing 100049, China*

[7]*Tsung-Dao Lee Institute, Shanghai 200240*

[8]*Collaborative Innovation Center of Quantum Matter, Beijing, 100190, China*

Here we report the observation of pressure-induced melting of antiferromagnetic (AFM) order and emergence of a new quantum state in the honeycomb-lattice halide α-RuCl₃, a candidate compound in the proximity of quantum spin liquid state. Our high-pressure heat capacity measurements demonstrate that the AFM order smoothly melts away at a critical pressure ($P_C$) of 0.7 GPa. Intriguingly, the AFM transition temperature displays an increase upon applying pressure below the $P_C$, in stark contrast to usual phase diagrams, for example in pressurized parent compounds of unconventional superconductors. Furthermore, in the high-pressure phase an unusual steady of magnetoresistence is observed. These observations suggest that the high-pressure phase is in an exotic gapped quantum state which is robust against pressure up to ~140 GPa.

The concept of the quantum spin liquid (QSL) was originally proposed by Phil Anderson in 1973 [1], describing a system of interacting quantum spins that does not order even at zero temperature [2-5]. At low-energy, such a system hosts an unusual feature that its propagating excitations possess only spin (no charge) degree of freedom, namely a spin liquid. Applying by this concept, he later proposed that the superconductivity in copper oxide superconductors can evolve from such a spin liquid state [6,7]. Although attempts to confirm such a state produced a null result in copper oxide superconductors, later the state in a pressurized organic conductor was observed [8], subsequently stimulating the physics community to further explore this exotic phenomenon [9-11].

Recent theoretical studies propose that the spin liquid state possesses long-range quantum entanglement [12] and sometimes non-trivial topological properties, making it a strong candidate for quantum computing applications [9]. Further developments have shown that the lattice structure hosting various types of frustrated couplings likely plays a central role in achieving the QSL state [11,13-17]. The realization of such a state in actual materials is of significant importance, however no solid evidence for the existence of such materials has been found in the laboratory despite a decades-long search. More recently, several experimental studies have been performed in the search for the candidate materials exhibiting properties similar to spin liquid ground states [3,4,10,18-25]. Among these, the iridium compounds $A_2IrO_3$ (A=Li, Na) [19,20] and α-$RuCl_3$ [3,4,24,25] with the honeycomb lattice were proposed as possible candidates, but, disappointingly, long-range zig-zag magnetic ordering was found instead [3,4,24]. Very recently it was reported that the magnetic order in α-RuCl$_3$ can be suppressed under external magnetic field around 7.5 T [26-31]. However, it was unclear whether the magnetic field-induced fluctuation favors the QSL correlations or not. In this study, we report results obtained from complementary high pressure measurements on α-RuCl$_3$.

Figure 1a and 1b show the temperature dependence of the magnetic ordering-related contribution to the heat capacity for the sample A, upon increasing pressure and releasing pressure respectively. The background thermal contribution to the heat capacity is removed to make the change induced by spin ordering more prominent. At a near-ambient pressure of 0.1 GPa, an anti-ferromagnetic (AFM) phase is observed with a transition temperature ($T_N$) ~7 K, consistent with results reported previously [3,4,24]. Intriguingly, unlike the behaviors, commonly seen in copper oxide and iron pnictide superconductors whose $T_N$ decreases upon increasing pressure, the $T_N$ of the α-RuCl$_3$ grows steadily under pressure until the magnetic order disappears at 0.7 GPa. The steady increase of $T_N$ indicates a stronger magnetic coupling and correspondingly a larger spin gap [4], given a shorter inter-atomic distance at higher pressure, all the way until the disappearance of the ordering. Note that our high-pressure x-ray diffraction measurements at two different synchrotron sources show no crystal structure phase transition up to 150 GPa, the highest measured pressure (Fig.S1 in Supplementary Information).

Figure 1c clarifies the nature of the demise of the magnetic order. The ordering-related entropy reduction, Δ$S$, obtained from integrating the heat capacity in Fig. 1a & 1b, reduces gradually to zero near 0.7 GPa. Upon decreasing the pressure below 0.7 GPa, $\Delta S$ grows smoothly again without any sign of hysteresis. This indicates that, the loss of the AFM order beyond 0.7 GPa is in fact a second-order quantum phase transition despite the first-order appearance of the phase diagram shown in Fig. 1d. Since the observed AFM melts away continuously with a *finite $T_N$*, these results manifest that beyond 0.7 GPa the system no longer host this AFM order even at zero temperature (or any other order, from our heat capacity data.)

A strikingly novel feature never seen in known systems is demonstrated in the phase diagram of Fig.1d: the classical critical transition at $T_N$ (marked in blue) is overwhelmed at *finite* temperature by quantum fluctuations (marked in red) at the critical pressure ($P_c$). In typical phase diagrams, the smaller order parameter near the quantum critical points can be depleted more easily at a lower temperature via, for example, thermal population of the low-energy Goldstone mode. This leads to a natural reduction of transition temperature smoothly to zero right at the quantum critical point, as illustrated in Fig. 1e. That is, the classical critical line smoothly connects to the quantum critical point. The case of α-RuCl$_3$, however, shows an exceptional increasing transition temperature when approaching the quantum critical point at $P_c$, where the order parameter decreases to zero (see illustration in Fig. 1f). This is conceptually possible if the AFM order is gapped [4] and a growing gap size can in principle delay the thermal depletion of the order to a higher temperature.

There is, however, a more profound generic reason for this novel separation of classical critical points at $T_N$ and the quantum critical point at $P_C$. Classically the ordering of the magnetic phase dictates a long-range correlation in the phase and near $T_N$. On the other hand, nearly all proposed QSL states are known to have negligible correlation beyond the nearest neighboring sites. Specially, the pure Kitaev QSL state relevant to α-RuCl$_3$ here, has identically zero correlation beyond first neighbors [32]. To switch to such an extremely short-range correlated state from a long-range correlated magnetic state, the divergence at the quantum critical point must be very singular, distinct from (and stronger than) the classical critical points [33]. It is thus very difficult, if even possible, to smoothly "connect" the classical critical points to the quantum critical point. From this consideration, the novel separation of the classical critical points from the quantum one and the overpowering of the latter against the former at *finite* temperature as seen here should be a rather generic feature of phase transition involving the QSL.

In other words, the melting of AFM at finite temperature in Fig. 1d indicates an "opposite" nature of the spin correlation between the normal paramagnetic states right above $T_N$ and the exotic quantum state right beyond $P_c$, a possible QSL with extremely short-range correlation.

This new quantum state above $P_c$ is quite stable under pressure as high as ~140 GPa (Fig. 2), and its nature was further verified by our resistivity measurements in a anisotropic hydrostatic environment (see Supplementary Information) within the diamond anvil cell under applied magnetic field up to 7 T. As shown in Fig. 2, the resistivity is surprisingly insensitive to the applied magnetic field. This lack of magnetoresistance effect, together with the signature of absence of AFM order at pressure above $P_c$, suggests that the pressure-induced new quantum state may be associated with a QSL state. It is noteworthy that our observation of the possible QSL behavior is in agreement with previous inelastic neutron scattering observations [3,4], in which a large continuum at low energy that resembles well the continuum of fractionalized particles pairs expected in the QSL is found, in addition to the regular gapped spin excitations. One thus expects that under a small pressure, when the ordered component of the system melts away, the remaining component of the ground state of α-RuCl$_3$ would demonstrates the same QSL-like continuum excitation. What the real ground state of α-RuCl$_3$ at pressure above 0.7 GPa is an open question which deserves further investigations by high pressure neutron studies.

More surprisingly, the dominance of the new quantum state appears to persist to the highest pressure (~140 GPa) investigated, even when a small number of charge carriers have been introduced to the system. Figure 3 shows the high pressure resistivity measurements for the sample B to the sample D which were in the quasi-hydrostatic (sample B) and anisotropic quasi-hydrostatic (sample C and D) pressure environments. It is seen that, at a fixed temperature, the resistivity of the samples consistently decreases upon increasing pressure, implying that the population of charge carriers is on the increase. Note that the resistivity of the α-RuCl$_3$ sample subjected to pressure as high as ~ 110 GPa still exhibits an insulating behavior (Fig.3h). We thus infer that the carriers scatter is more strongly at low temperature and impact on the electron correlation. Future theoretical studies on properties of such a new quantum state would be extremely valuable to illuminate the physics of charge carriers behind our observation.

Fitting our temperature dependent resistivity to *exp* ($\varepsilon_A/2k_BT$) to extract the "activation" energy $\varepsilon_A$ (see Supplementary Information), we found three distinct regimes of charge transport below 27 GPa where the charge gap exists and the system is truly an insulator, as shown in Fig. 4. Note that the $\varepsilon_A$ obtained in this study should be taken as a mirror of the interaction energy of electron spin ($\varepsilon_S$) and charge ($\varepsilon_C$), instead of the simple activation energy as adopted in semiconductor physics. At pressures below ~3.4 GPa, $\varepsilon_A$ grows steadily with the increase of $T_N$ under pressure. It is thus reasonable to believe that the low-energy barrier for the carriers to overcome is the energy scale of the short-range spin correlation, $\varepsilon_S$. As the bands grow wider along with the reduced inter-atomic distance at higher pressure, the charge gap, $\varepsilon_C$, is expected to reduce monotonically, eventually becoming lower than $\varepsilon_S$ and dominating the thermal activation of charge transport. This explains the sudden drop of $\varepsilon_A$ around 10 to 30 GPa. We have conducted a density functional theory calculation and found that the charge gap should close at around 20-30 GPa (Fig.S3 in Supplementary Information). Furthermore, Figure 4b shows that the absolute value of the resistivity drops by more than five orders of magnitude at high pressure, now well beyond the regime of a well gapped charge system.

In conclusion, our results show that a small pressure can completely suppress the AFM ordered component and drive the α-RuCl$_3$ into a new quantum state. This exotic state is robust against the pressure up to 140 GPa. These results allow us to obtain a novel phase diagram involving two states with different nature, on the crossover of which the pressure-enhanced quantum fluctuation destroys the thermally-driven Neél critical transition in the ambient-pressure phase at *finite* temperature.

**Acknowledgements**

We thank Tao Xiang, Yi Zhou, Gang Chen, Zhongyi Liu, Weiqiang Yu, Jingsheng Wen and Jianxin Li for helpful discussions. The work in China was supported by the National Key Research and Development Program of China (Grant No. 2017YFA0302900, 2016YFA0300300 and 2017YFA0303103), the NSF of China (Grants No. 91321207, No. 11427805, No. 11404384, No. U1532267, No. 11604376, No.11674220 and No.11447601), the Strategic Priority Research Program (B) of the Chinese Academy of Sciences (Grant No. XDB07020300) and Ministry of Science and Technology (Grants No. 2016YFA0300500 and No.2016YFA0300501). The work at Princeton was supported by the Gordon and Betty Moore Foundation EPiQS initiative, grant GBMF-4412.

†Corresponding authors

L.S. (llsun@iphy.ac.cn) or W.K. (weiku@mailaps. org).

* These authors contributed equally.

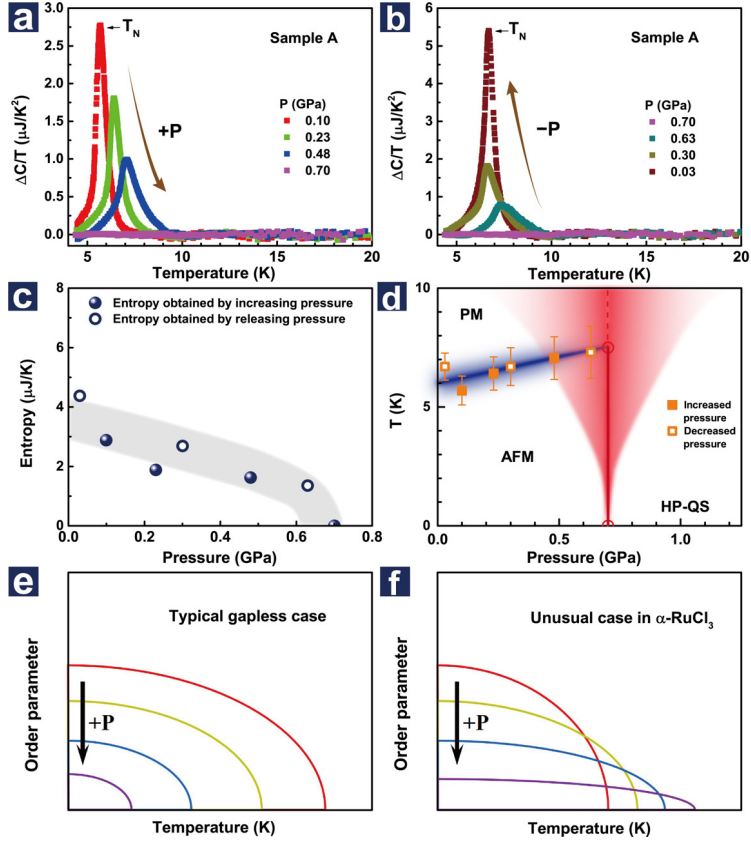

FIG. 1. Results obtained from high pressure heat capacity measurements for α-RuCl$_3$. (a) and (b) Temperature dependence of the heat capacity (in the form of ΔC/T) for the sample A upon increasing pressure and releasing pressure respectively. (c) Pressure dependence of magnetic entropy integrated from the data of Fig.1a and 1b. (d) Pressure-Neel temperature phase diagram for α-RuCl$_3$. $T_N$ represents Neel temperature, AFM and PM stand for antiferromagnetic order and paramagnetic states respectively and HP-QS represents high-pressure new quantum state. The red domain indicates the quantum fluctuation regime. (e) Typical temperature dependence of the order parameter for systems without a spin gap when approaching a quantum critical point via pressure. (f) Unusual case reported here with a larger spin gap when approaching the quantum critical point: smaller zero-temperature order parameter that survives higher $T_N$.

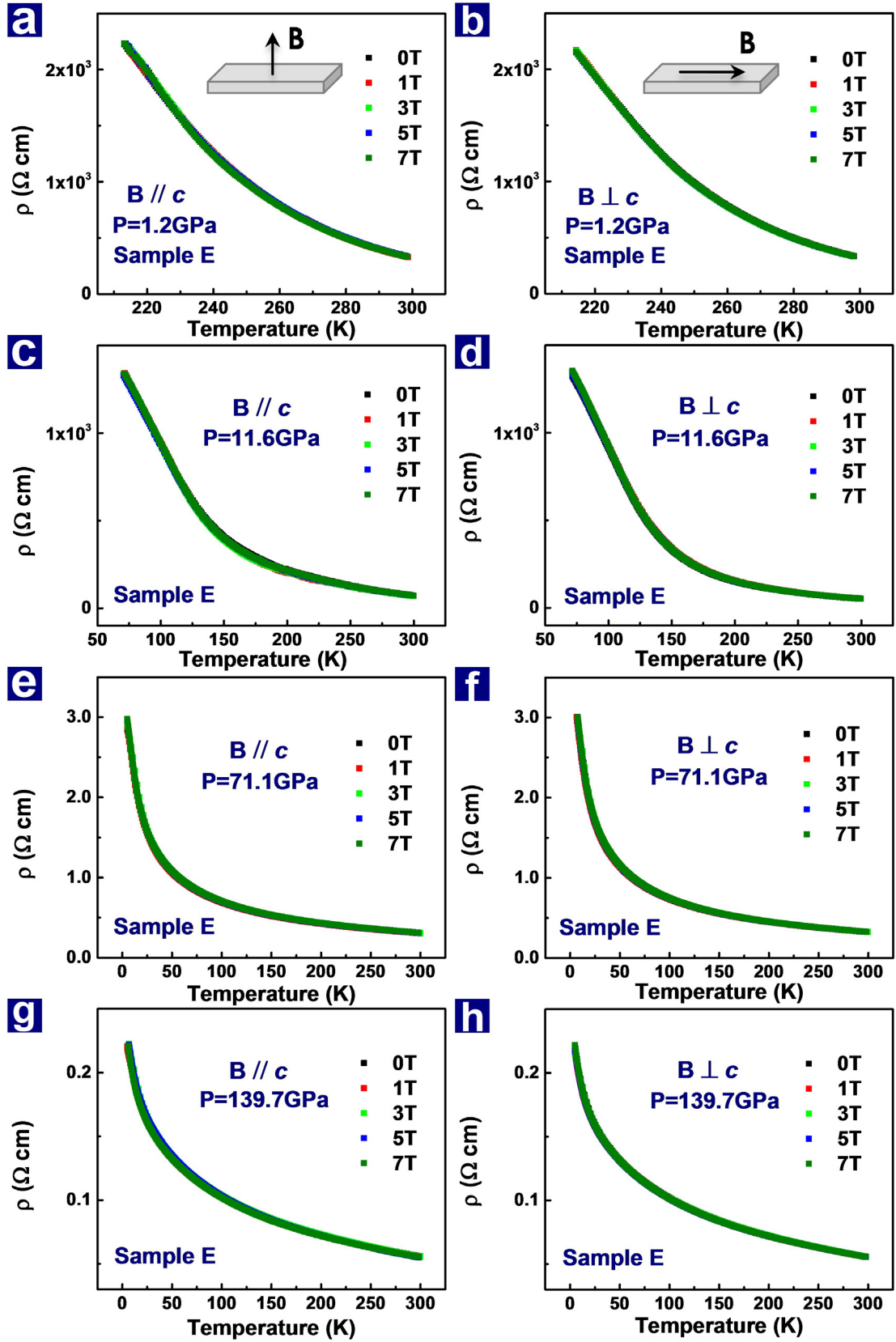

FIG. 2. Electrical resistivity results obtained from the measurements under different

magnetic fields for pressurized α-RuCl$_3$. (a) (c) (e) and (g) Temperature dependence of resistivity under the magnetic field parallel to the *c*-axis of the honeycomb lattice. (b) (d) (f) and (h) Temperature dependence of resistivity under the magnetic field perpendicular to the *c*-axis of the honeycomb lattice.

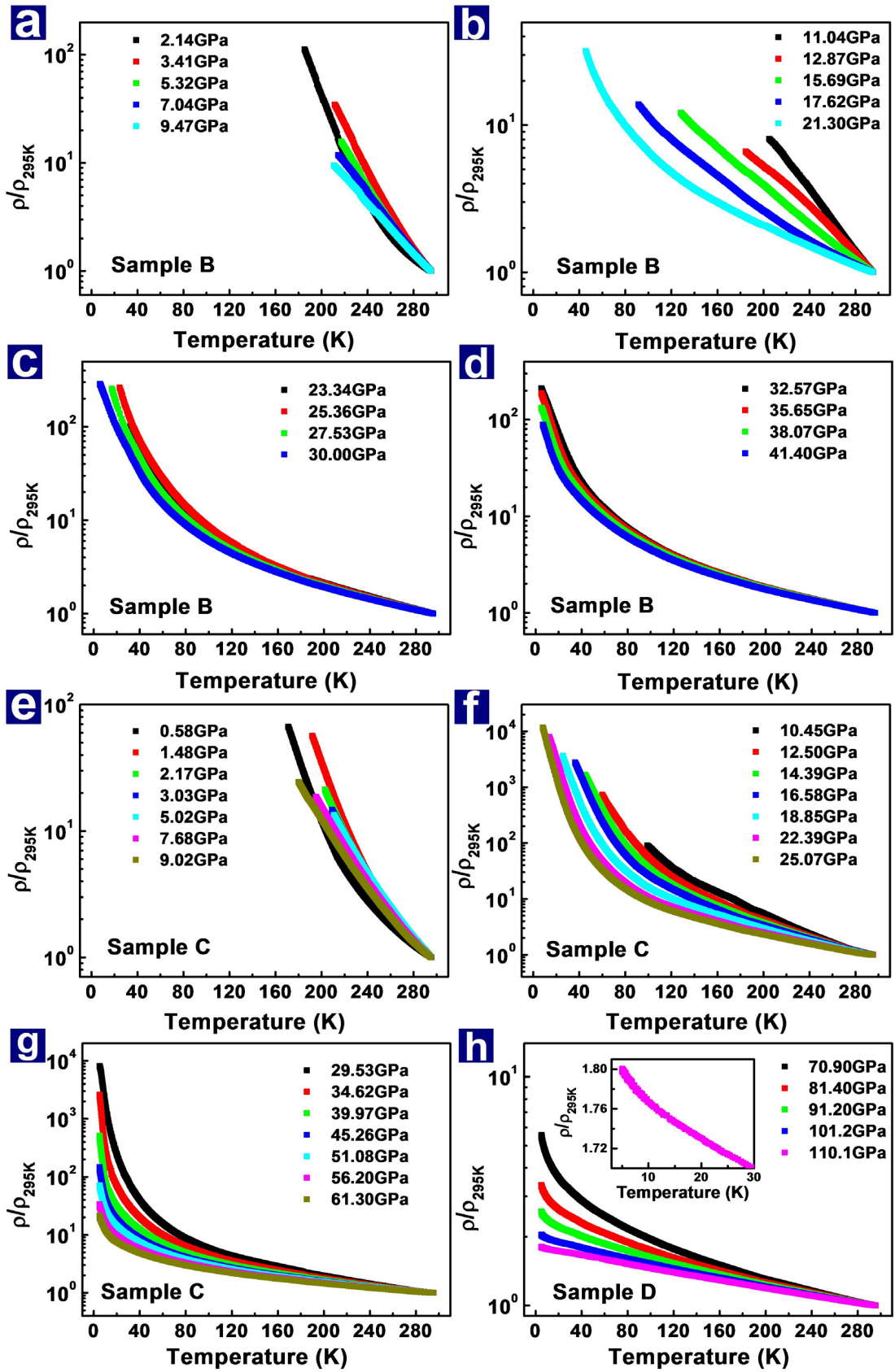

FIG. 3. Electrical resistivity as a function of temperature at different pressures for α-RuCl$_3$. Figure (a) to (d) Temperature dependence of electrical resistivity for the sample B obtained at quasi-hydrostatic pressures. Figure (e) to (h) Temperature dependence of electrical resistivity for the sample C and the sample D obtained at anisotropic quasi-hydrostatic pressures. The inset displays an enlarge view of the resistance-temperature curve obtained at ~110 GPa.

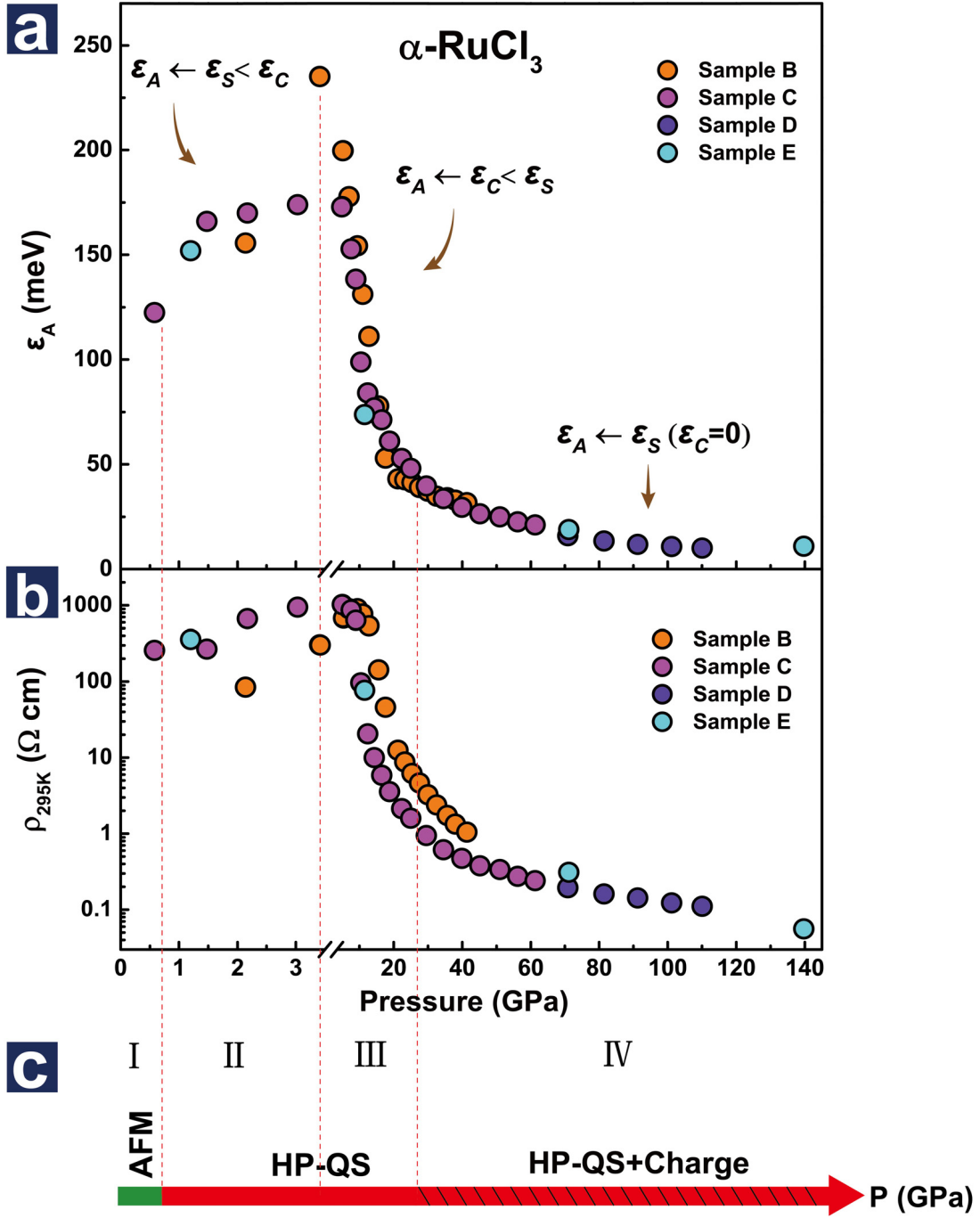

FIG. 4. Summary of high-pressure heat capacity and electrical resistivity for α-RuCl$_3$. (a) Pressure dependence of energy gap ($\varepsilon_A$). (b) Resistivity measured at 295 K as a function of pressure. (c) Evaluation of ground states with pressure. $\varepsilon_A$ is the "activation" energy, $\varepsilon_S$ and $\varepsilon_C$ stand for the interaction energy of spin and charge. HP-QS represents high-pressure new quantum state.

17